\documentstyle[preprint,pra,aps,epsf]{revtex}

\begin{document}
\draft
\preprint{CfA preprint 4659}
\title{Long-range interactions of metastable
helium atoms}
\author{Zong-Chao Yan and J. F. Babb}
\address{
Institute for Theoretical Atomic and Molecular Physics,\\
Harvard-Smithsonian Center for Astrophysics,\\ 
60 Garden Street, Cambridge, MA 02138 }

\maketitle
\begin{abstract}
Polarizabilities, dispersion coefficients, and long-range atom-surface
interaction potentials are calculated for the $n=2$ triplet and
singlet states of helium using highly accurate, variationally
determined, wave functions.
\end{abstract}
\narrowtext

The advent of doubled basis sets has made it possible to calculate
precisely many properties of two-electron atomic
systems~\cite{Drake-LRF,DraYan92,DraYan94,Drake-handbook}.  We apply
variational methods developed previously and demonstrated  for the
helium atom~\cite{YanBabDal96} to calculate nonrelativistic values of
the electric dipole, quadrupole, and octupole polarizabilities and
corresponding dispersion coefficients for the metastable $n=2$ singlet
and triplet states, respectively, He$(2\,^1\!S)$ and
He$(2\,^3\!S)$. Additionally,  potentials for the atom-wall interaction
of a He$(2\,^1\!S)$ or a He$(2\,^3\!S)$ atom and a single perfectly
conducting wall or a dielectric wall are calculated with the
inclusion of retardation effects due to the finite speed of light.
Our results for atom-wall interactions are germane to
experiments involving atom-evanescent wave mirrors~\cite{LanCouLab96}.

In this paper the notation of Ref.~\cite{YanDalBab97} is followed very
closely; references to equations of Ref.~\cite{YanDalBab97} will be
preceded by the symbol~I.  Atomic units are used throughout.

The dispersion interaction of two like atoms can be written
\begin{equation}
\label{disp}
U (R) = - C_6R^{-6} - C_8R^{-8} - C_{10}R^{-10},
\end{equation}
where the coefficients $C_6$, $C_8$, and $C_{10}$ are the van der
Waals coefficients, $R$ is the interatomic distance, and
\begin{equation}
\label{C6}
C_6 = (3/\pi)G(1,1)\,,
\end{equation}
\begin{equation}
\label{C8}
C_8 =  (15/\pi)G(1,2)\,,
\end{equation}
\begin{equation}
\label{C10}
C_{10} =  (28/\pi)G(1,3)+ (35/\pi)G (2,2)\,,
\end{equation}
with 
\begin{equation}
G(l,m) = \int_{0}^{\infty}
\alpha_{l}(i\omega)
\alpha_{m}(i\omega)d\omega\, ,
\end{equation}
where $\alpha_{l}(i\omega)$ is the $2^{l}$-pole dynamic
polarizability function evaluated at imaginary frequency defined by
Eqs.~(6)--(9) of Ref.~\cite{YanBabDal96}, and similarly for
$\alpha_{m}(i\omega)$.

When the effects of retardation due to the finite speed
of light are considered the potential $U(R)$, Eq.~(\ref{disp}),
can be replaced by~\cite{CasPol48,PowThi96}
\begin{equation}
\label{retpot}
V (R) = - C_6f_6 (R)R^{-6}- C_8f_8 (R)R^{-8}-C_{10}f_{10} (R)R^{-10},
\end{equation}
The coefficient $f_{10}(R)$ will not be considered in this paper as 
the $C_{10}$ term 
is usually negligible.  Expressions for the retardation coefficients,
$f_6 (R)$ and $f_8 (R)$, as integrals involving the dynamic electric
dipole polarizabilities, are given in Eqs.~I-(5) and I-(7).

The form (\ref{retpot}) intrinsically includes certain relativistic
effects, so that when $f_6 (R)$ and $f_8 (R)$ are expanded in powers
of the fine structure constant $\alpha_{\rm fs}=1/137.035\,989\,5$ for
small distances
\begin{equation}
\label{atom-small}
V (R )  \sim -R^{-6}[C_6 - \alpha_{\rm fs}^2R^2 W_4 ]
       -R^{-8}[C_8 - \alpha_{\rm fs}^2R^2 W_6 ] ,
\end{equation}
where 
\begin{equation}
\label{W4}
W_4 =  \frac{1}{\pi} \int_0^\infty 
       d\omega\;\omega^2 \alpha_1^2 (i\omega)
\end{equation}
and 
\begin{equation}
\label{W6}
W_6 = \frac{3}{\pi} \int_0^\infty 
       d\omega\;\omega^2 \alpha_1 (i\omega)\alpha_2 (i\omega) .
\end{equation}
The relativistic origin of the coefficient $W_4$
has been discussed by Power and Zienau~\cite{PowZie57},
see also~\cite{MeaHir66}.
The coefficient $W_6$ of the factor  $\alpha_{\rm fs}^2/R^6$ in
(\ref{atom-small}) corresponds to the theory of Power and
Thirunamachandran~\cite{PowThi96} and is equal to the coefficient
$W_{LL;4,2}$ in the theory of Meath and Hirschfelder
based on the Breit-Pauli Hamiltonian~\cite{MeaHir66}.
As the distance increases retardation arising from the finite speed
of light becomes important and the potential approaches its asymptotic
form, see Eqs.~I-(13) and I-(14),
\begin{equation}
\label{asymp-atom}
V(R) \sim -K_7R^{-7} -K_9R^{-9} ,
\end{equation}
with
\begin{equation}
\label{asymp-coeffs}
K_7 = \frac{23}{4\pi} \frac{\alpha_1^2 (0)}{\alpha_{\rm fs}}
 = 250.81\; \alpha_1^2 (0) , \qquad 
K_9 = \frac{531}{16\pi} \frac{\alpha_1 (0)\alpha_2 (0)}{\alpha_{\rm fs}}
 = 1\,447.6\; \alpha_1 (0)\alpha_2 (0).
\end{equation}

An expression for the potential $V_{{\rm At}D} (R,\epsilon)$ for the
interaction~\cite{DzyLifPit61,TikSpr93b} of an atom and a dielectric
wall was presented in Eq.~I-(15), where $R$ is the atom-wall distance
and $\epsilon$ is the dielectric constant of the wall. The expression
is a double integral that can be evaluated with knowledge of the
function $\alpha_1 (i\omega)$.  For small distances $V_{{\rm At}D}
(R,\epsilon)$ has the limiting form
\begin{equation}
\label{small-R}
V_{{\rm At}D} (R,\epsilon)  \sim 
     -\frac{C_3}{R^3}\frac{\epsilon-1}{\epsilon+1}, 
\end{equation}
where 
\begin{equation}
\label{C3-integral}
C_3 = \frac{1}{4\pi} \int_0^\infty d\omega  \alpha_1 (i\omega) .
\end{equation}
As the separation increases retardation becomes important and the
potential approaches its asymptotic form,
\begin{equation}
\label{dzyal}
V_{{\rm At}D} (R,\epsilon) \sim
     -\frac{K_4}{R^4} 
         \frac{\epsilon -1}{\epsilon +1} \phi (\epsilon),
\end{equation}
where 
$\phi (\epsilon)$ is given in Eq.~I-(21) and 
\begin{equation}
\label{casimir}
K_4 =  3 \alpha_1 (0)/  (8\pi\alpha_{\rm fs}) = 16.357\, \alpha_1 (0).  
\end{equation}

For a perfectly conducting wall $V_{{\rm At}D} (R,\epsilon)$ 
reduces to $V_{{\rm At}M} (R)$, where
\begin{equation}
\label{AtM}
V_{{\rm At}M} (R) \equiv  
V_{{\rm At}D} (R,\infty) 
     = -C_3f_3 (R)R^{-3} 
\end{equation}
and the retardation coefficient $f_3 (R)$ is an integral
involving $\alpha_1 (i\omega)$ and is given in Eq.~I-(26).  For small
distances $V_{{\rm At}M} (R) \sim -C_3/R^3$ and for asymptotically
large distances $V_{{\rm At}M} (R) \sim -K_4/R^4$.  Table~II of
Ref.~\cite{ZhoSpr95} summarizes  the various limits of $V_{{\rm
At}D} (R,\epsilon)$.

It has been shown that double basis sets work well for calculations
involving $S$ states of helium~\cite{DraYan94}.  
The basis set used here was  constructed
as in Ref.~\cite{YanBabDal96}
with basis set functions expressed using Hylleraas coordinates 
\begin{eqnarray}
\{\chi_{ijk} (\alpha,\beta) &=& 
r_1^i\,r_2^j\,r_{12}^k\,e^{-\alpha r_1-\beta r_2}\}\,.
\label{basis}
\end{eqnarray}
The explicit form for the wave function is
\begin{eqnarray}
\Psi({\rm\bf r}_1,{\rm\bf r}_2)& =& \sum_{ijk}\,[a_{ijk}^{(1)}\,
\chi_{ijk}(\alpha_1,\beta_1)
+a_{ijk}^{(2)}\,\chi_{ijk}(\alpha_2,\beta_2)]
\pm\, {\rm exchange}\,,
\label{wf}
\end{eqnarray}
and $i+j+k\le\Omega$. 
The convergence of the eigenvalues
is studied as $\Omega$ is progressively enlarged. Finally,
a complete optimization is performed with
respect to the two sets of nonlinear parameters $\alpha_1$,
$\beta_1$, and $\alpha_2$, $\beta_2$ by first calculating the
derivatives analytically in 
\begin{eqnarray}
{\partial E}\over {\partial \alpha} &=&2\bigg\langle\Psi\bigg|H\bigg|
{{\partial\Psi}\over {\partial\alpha}}\bigg\rangle
-2E\bigg\langle\Psi\bigg|
{{\partial\Psi}\over {\partial\alpha}}\bigg\rangle\,,
\label {eq:aa1}
\end{eqnarray}
where $\alpha$ represents any nonlinear parameter, 
$E$ is the trial energy, $H$ is the Hamiltonian, and
$\langle \Psi|\Psi \rangle=1$ is assumed, and then 
locating the zeros of the derivatives by Newton's method.
These techniques yield much improved convergence
relative to single basis set calculations.
The method of the evaluation of the two-electron integrals
in Hylleraas coordinates can be found in Ref. \cite{YanDra96}.

\begin{figure}[p]
\epsfxsize=1.\textwidth \epsfbox{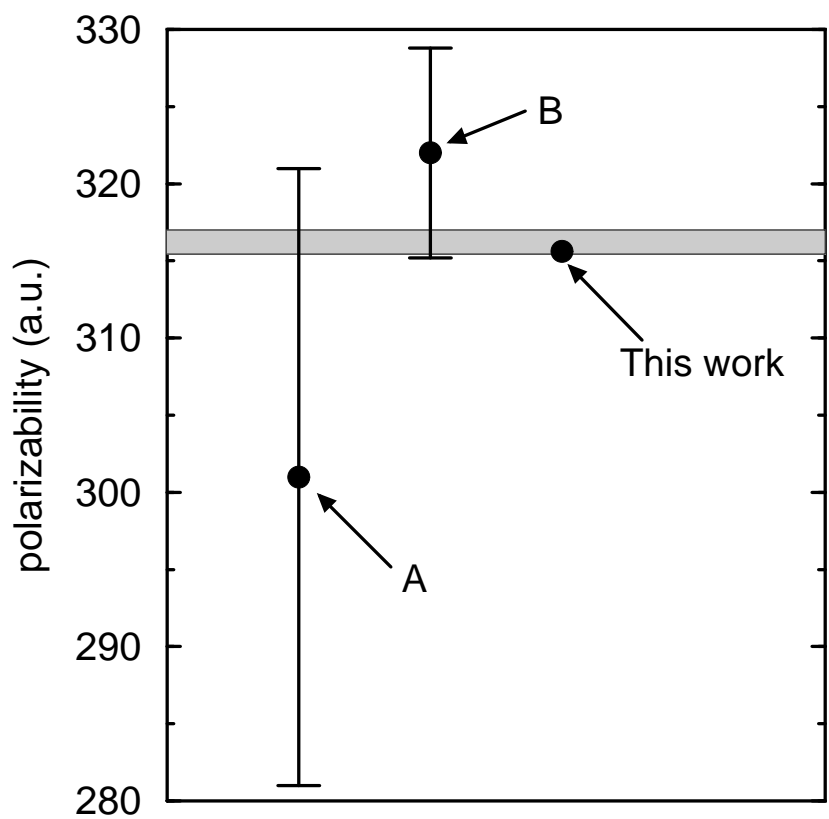}
\caption{Comparison of measured
values (A, B) and the upper and lower bounds of Glover and Weinhold
(shaded region)~\protect\cite{GloWei77b} with the present calculation of the
static polarizability for He($2\,^3\!S$). The point A is the
measurement of 
Crosby and Zorn~\protect\cite{CroZor77} and the
point B is that of 
Ekstrom {\it et al.\/}~\protect\cite{EksSchCha95}
determined in combination with measurements from
Molof {\it et al.\/}~\protect\cite{MolSchMil74}.
\label{triplet-pol-fig}}
\end{figure}

The expressions for the dynamic dipole polarizabilities, Eqs.~(6)--(9)
of Ref.~\cite{YanBabDal96}, were evaluated using the wave functions
determined by the variational method.  Values of the static
polarizabilities are given in Table~\ref{pol-this} for He($2\,^1\!S$)
and He($2\,^3\!S$).  The polarizabilities given in
Table~\ref{pol-this} are extrapolated results, with the convergence
studied as in Refs.~\cite{DraYan94} and \cite{YanBabDal96}, and the
estimated extrapolation error in the last digit is given in
parentheses with the listed values.  The largest basis set sizes used
consisted of 616 functions for the $S$ states, 910 functions for the
$P$ states, $931$ functions for the $D$ states, and 1092 functions for
the $F$ states.  The converged results are compared with some previous
calculations and experiments in Table~\ref{singlet-pol-compare} and
\ref{triplet-pol-compare}.  Ekstrom {\it et al.\/}~\cite{EksSchCha95}
determined the He($2\,^3\!S$) polarizability by combining their
measured Na polarizability with the Molof {\it et
al.\/}~\cite{MolSchMil74} measurement of the Na polarizability
relative to the polarizability of He($2\,^3\!S$).  For the triplet
state the experimental values of Ref.~\cite{CroZor77} and of
Refs.~\cite{EksSchCha95,MolSchMil74} and the bounds of Glover and
Weinhold are compared with our calculated polarizability in
Fig.~\ref{triplet-pol-fig}.

\begin{figure}[p]
\epsfxsize=1.\textwidth \epsfbox{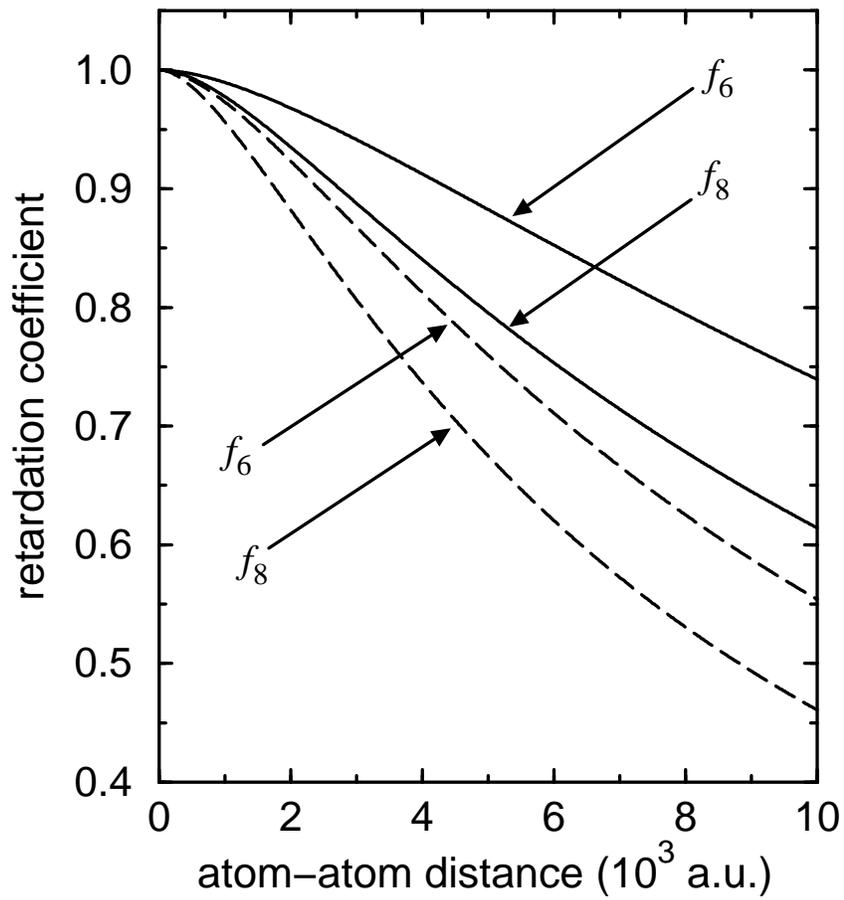}
\caption{Dimensionless retardation
coefficients for He($2\,^1\!S$) (solid line) 
and He($2\,^3\!S$) (dashed line).
\label{ret-coeffs-fig}}
\end{figure}

The dynamic polarizability functions were constructed
using the largest basis sets of each symmetry and used to evaluate the
atom-atom dispersion constants and retardation coefficients. Our
results for the dispersion constants are given in Table~\ref{disp-us},
with the estimated convergence errors given in parentheses, and the
results are compared to other calculations in
Tables~\ref{singlet-disp} and \ref{triplet-disp}.  The retardation
coefficients are given in Table~\ref{f-table} and
Fig.~\ref{ret-coeffs-fig}.

Chen and Chung~\cite{CheChu96} calculated the coefficients $W_4$ and
$W_6$ for He($2\,^1\!S$) and their results are compared with ours in
Table~\ref{rel-table}; their published value of $W_6$ was multiplied
by the factor $\frac{3}{2}$ to correspond to the theory of
Ref.~\cite{PowThi96} and the agreement is very good.

For the atom-wall interactions the values of the coefficients  $C_3$
can be obtained from the alternate expression
\begin{equation}
\label{C3-direct}
C_3 = \frac{1}{12} 
   \left \langle 0 \left| \left(\sum_{i=1}^N{\bf r}_i \right)^2
          \right| 0 \right\rangle ,
\end{equation}
which follows from integration of Eq.~(\ref{C3-integral}), where $N$ is
the number of electrons and $| 0 \rangle$ is accordingly the
$2\,^1\!S$ or the $2\,^3\!S$ wave function.  Since high-precision
matrix elements are available~\cite{Drake-handbook,Drake-private} 
Eq.~(\ref{C3-direct}) was used 
to obtain the coefficients $C_3 (2\,^1\!S) =2.671\,212\,717\,025$ and $C_3
(2\,^3\!S) =1.900\,924\,084\,097$.

\begin{figure}[p]
\epsfxsize=1.\textwidth \epsfbox{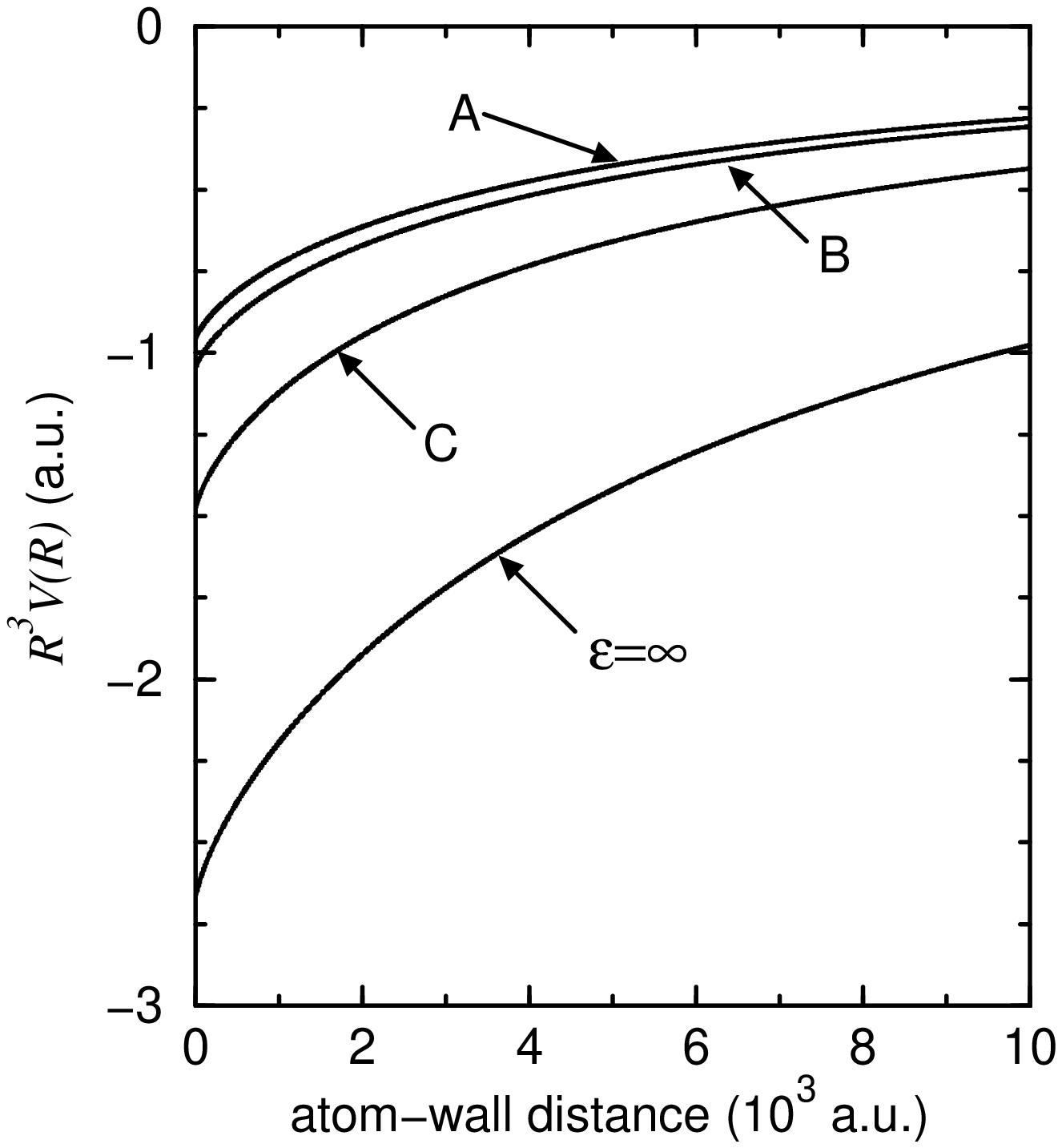}
\caption{Potentials $R^3 V(R)$ for He($2\,^1\!S$) atom-wall interactions. The
labels A, B, and C correspond, respectively,
to dielectric constants $\epsilon$ of 2.123, 2.295, 
and 3.493.
\label{aw-singlet-fig}}
\end{figure}

\begin{figure}[p]
\epsfxsize=1.\textwidth \epsfbox{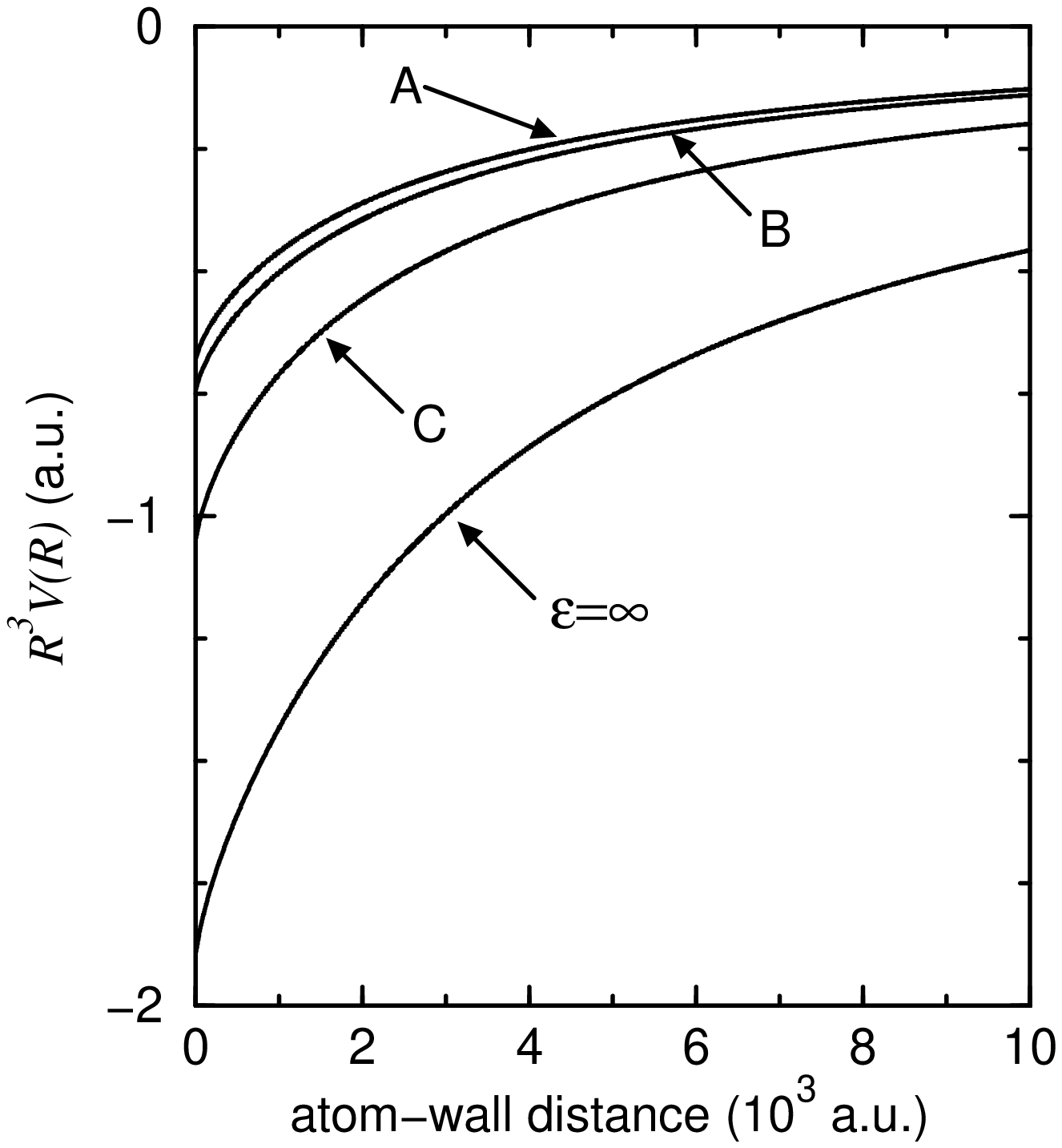}
\caption{Potentials $R^3 V(R)$  for He($2\,^3\!S$) atom-wall interactions. The
labels A, B, and C correspond
respectively, to dielectric constants $\epsilon$ of 2.123, 2.295, 
and 3.493.
\label{aw-triplet-fig}}
\end{figure}

The dynamic dipole polarizability was used to evaluate the potential
for various dielectric walls.  Results for He$( 2\,^1\!S)$ are given
in Table~\ref{glass-table-singlet} and Fig.~\ref{aw-singlet-fig} and
those for He$( 2\,^3\!S)$ are given in Table~\ref{glass-table-triplet}
and Fig.~\ref{aw-triplet-fig}.  The dielectric materials represented
in the tables correspond to fused silica ($\epsilon=2.123$), BK-7
glass ($\epsilon=2.295$), and a GaAs-type material
($\epsilon=3.493$). The tabulated potentials may be helpful in
planning and analyzing experiments with atom-evanescent wave mirrors,
see for example Ref.~\cite{LanCouLab96}.

We thank Professor G.~W.~F. Drake and 
Dr. P.~L. Bouyer for helpful communications.
The Institute for Theoretical Atomic and Molecular Physics is
supported by a grant from the National Science Foundation to the
Smithsonian Institution and Harvard University. 
ZCY was also supported by the Natural Sciences and Engineering
Research Council of Canada.

\begin{table}
\caption{Values of the static polarizabilities $\alpha_1(0)$,
$\alpha_2(0)$, and $\alpha_3(0)$ for
the $2\,^1\!S$ and $2\,^3\!S$ states of He. Numbers
in parentheses represent the estimated error
in the last digit of 
the listed, extrapolated value.}
\label{pol-this}
\begin{tabular}{l c c c}
\multicolumn{1}{l}{State}&
\multicolumn{1}{c}{$\alpha_1(0)$}&
\multicolumn{1}{c}{$\alpha_2(0)$}&
\multicolumn{1}{c}{$\alpha_3(0)$}\\
\tableline
$2{}^1S$ &  800.316\,33(7)   & 7\,106.053\,7(5) & 293\,703.50(6) \\
$2{}^3S$ &  315.631\,468(12) & 2\,707.877\,3(3) & 88\,377.325\,3(7)\\
\end{tabular}
\end{table}
\begin{table}
\caption{Comparison of static multipole polarizabilities
$\alpha_1(0)$, $\alpha_2(0)$, and $\alpha_3(0)$
for He($2\,^1\!S$). For the experimental
value numbers in parenthesis give the quoted error.
\label{singlet-pol-compare}}
\begin{tabular}{l c r@{}l r@{}l r@{}l}
\multicolumn{1}{l}{Author (year)}&
\multicolumn{1}{c}{Ref.}&
\multicolumn{2}{c}{$\alpha_1(0)$}&
\multicolumn{2}{c}{$\alpha_2(0)$}&
\multicolumn{2}{c}{$\alpha_3(0)$}\\
\tableline
Crosby and Zorn (77) Expt.& \cite{CroZor77} &
 729&(88)& && &\\
Chung and Hurst (66) & \cite{ChuHur66} &
 801&.9& && &\\
Drake (72)  & \cite{Dra72} &
 800&.2& && &\\
Chung (77) & \cite{Chu77}  &
 801&.10& && &\\
Glover and Weinhold (77)& \cite{GloWei77a}  &
 803&.31$\pm$6.61\tablenote{Bounded theoretical value.}& && &\\
Lamm and Szabo (80), ECA  & \cite{LamSza80} &
 790&.8& && & \\
R\'{e}rat {\it et al.\/} (93) & \cite{RerCafPou93} &
 803&.25& 6870&.9 & &\\
Chen (95) & \cite{Che95a} &
 800&.34 & && &\\
This work &  & 
 800&.316\,33(7)    & 7\,106&.053\,7(5) & 293\,703&.50(6)  \\
\end{tabular}
\end{table}

\begin{table}
\caption{Comparison of static multipole polarizabilities
$\alpha_1(0)$, $\alpha_2(0)$, and $\alpha_3(0)$
for He($2\,^3\!S$).\label{triplet-pol-compare}}
\begin{tabular}{l c r@{}l r@{}l r@{}l}
\multicolumn{1}{l}{Author (year)}&
\multicolumn{1}{c}{Ref.}&
\multicolumn{2}{c}{$\alpha_1(0)$}&
\multicolumn{2}{c}{$\alpha_2(0)$}&
\multicolumn{2}{c}{$\alpha_3(0)$}\\
\tableline
Crosby and Zorn (77) Expt.& \cite{CroZor77} &
301&(20)& &&  &\\
Ekstrom {\it et al.\/} (95) Expt.& \cite{EksSchCha95,MolSchMil74} &
322&(6.8)& &&  &\\
Bishop and Pipin (93) & \cite{BisPip93} &
315&.631 & 2\,707&.85 &  88\,377&.2 \\
R\'{e}rat {\it et al.\/} (93) & \cite{RerCafPou93} &
315&.92 & 2\,662&.02 & & \\
Glover and Weinhold (77) & \cite{GloWei77a} &
316&.24$\pm$0.78\tablenote{Bounded theoretical value.} & &&  & \\
Drake (72)  & \cite{Dra72} &
 315&.608& && &\\
Chung (77) & \cite{Chu77}  &
315&.63 & && & \\
Chen and Chung (96), $B$ Spline & \cite{CheChu96} & 
315&.63 & 2\,707&.89 & 88\,377&.4 \\
Chen and Chung (96), Slater & \cite{CheChu96} & 
315&.611 & 2\,707&.81 & 88\,356&.2 \\
Chung and Hurst (66) & \cite{ChuHur66} &
315&.63 & && & \\
Chen (95) & \cite{Che95a} &
315&.633 & && & \\
This work &  & 
 315&.631\,468(12)  & 2\,707&.877\,3(3) &  88\,377&.325\,3(7)\\
\end{tabular}
\end{table}

\begin{table}
\caption{Values of $C_6$, $C_8$, and $C_{10}$ for the interaction of
two He atoms.}
\label{disp-us}
\begin{tabular}{l c c c}
\multicolumn{1}{l}{System}&
\multicolumn{1}{c}{$C_6$}&
\multicolumn{1}{c}{$C_8$}&
\multicolumn{1}{c}{$C_{10}$}\\
\tableline
$2\,^1\!S$-$2\,^1\!S$ & 
   11\,241.052(5)    & 817\,250.5(4)    & 108\,167\,630(54) \\
$2\,^3\!S$-$2\,^3\!S$ & 
   3\,276.680\,0(3)  & 210\,566.55(6)   & 21\,786\,760(5)  \\
\end{tabular}
\end{table}

\begin{table}
\caption{Comparison of $C_6$, $C_8$, and $C_{10}$
for the He($2\,^1\!S$)-He($2\,^1\!S$) system.}
\label{singlet-disp}
\begin{tabular}{l c r@{}l r@{}l r@{}l}
\multicolumn{1}{l}{Author (year)}&
\multicolumn{1}{c}{Ref.}&
\multicolumn{2}{c}{$C_6$}&
\multicolumn{2}{c}{$C_8$}&
\multicolumn{2}{c}{$C_{10}$}\\
\tableline
Glover and Weinhold (77)  
       & \cite{GloWei77b}& 11\,330&$\pm$630\tablenote{Bounded theoretical value.}
               && &&\\
R\'{e}rat {\it et al.\/} (93) & \cite{RerCafPou93} &
11\,360&        & 812\,500& && \\
Victor {et al.\/} (68)     & \cite{VicDalTay68}&
11\,300&        & && & \\
Lamm and Szabo (80), ECA  & \cite{LamSza80} &
10\,980&        & && & \\
Chen (95)                 & \cite{Che95b} & 
11\,244&        & 817\,360&   & 108\,184\,000& \\
This work                &                 &
11\,241&.052(5) & 817\,250&.5(4)    & 108\,167\,630&(54) \\
\end{tabular}
\end{table}

\begin{table}
\caption{Comparison of $C_6$, $C_8$, and $C_{10}$
for the He($2\,^3\!S$)-He($2\,^3\!S$) system.}
\label{triplet-disp}
\begin{tabular}{l c r@{}l r@{}l r@{}l}
\multicolumn{1}{l}{Author (year)}&
\multicolumn{1}{c}{Ref.}&
\multicolumn{2}{c}{$C_6$}&
\multicolumn{2}{c}{$C_8$}&
\multicolumn{2}{c}{$C_{10}$}\\
\tableline
Glover and Weinhold (77)  &
    \cite{GloWei77b}& 3\,289&$\pm$90\tablenote{Bounded theoretical value.} && &&\\
Victor {et al.\/} (68)     & \cite{VicDalTay68}&
3\,290&        & && & \\
Lamm and Szabo (80), ECA  & \cite{LamSza80} &
3\,300&        & && & \\
R\'{e}rat {\it et al.\/} (93) & \cite{RerCafPou93} &
3\,279&        & 208\,600& && \\
Bishop and Pipin (93)     & \cite{BisPip93}&
3\,276&.677\,0 &  210\,563&.99 &   21\,786\,484&\\
Chen (95)                 & \cite{Che95b} & 
3\,276&.1      & 210\,520&   & 21\,783\,800& \\
Chen and Chung (96), $B$ spline       & \cite{CheChu96}&
3\,276&.10     & 210\,518&   & 21\,783\,800& \\
Chen and Chung (96), Slater    & \cite{CheChu96}&
3\,275&.90     & 210\,507&   & 21\,780\,200& \\
This work                &                 &
3\,276&.680\,0(3) & 210\,566&.55(6)   & 21\,786\,760&(5) \\
\end{tabular}
\end{table}

\begin{table}
\begin{center}
\caption{
The dimensionless retardation coefficients $f_6 (R)$ and $f_8 (R)$ for
the atom-atom interaction.  The dispersion coefficients $C_6$ and
$C_8$ are also given. In the last line, labeled ``Asymptotic''
the values calculated using the asymptotic forms $f_6 \sim K_7/(RC_6)$
and $f_8 \sim K_9/(RC_8)$ are given in, respectively
cols. 2,4 and cols. 3,5, with $K_7$ and $K_9$ given
in Eq.~(\protect\ref{asymp-coeffs}).}
\label{f-table}
\begin{tabular}{lcccc}
 \multicolumn{1}{c}{ }&\multicolumn{2}{c}{He($2\,^1\!S$)-He($2\,^1\!S$)}&
  \multicolumn{2}{c}{He($2\,^3\!S$)-He($2\,^3\!S$)} \\
\cline{2-3}\cline{4-5}
     &  \multicolumn{1}{c}{$C_6$} &\multicolumn{1}{c}{$C_8$} &
     \multicolumn{1}{c}{$C_6$} &\multicolumn{1}{c}{$C_8$} \\
    &  112\,41.052(5)   & 817\,250.5(4)  &
     3\,276.680\,0(3) & 210\,566.55(6) \\
    $R$ & $f_6 (R)$ & $f_8 (R)$ & $f_6 (R)$ & $f_8 (R)$ \\
\hline
     10 & 0.999998 & 0.999996 & 0.999995 & 0.999992 \\  
     15 & 0.999996 & 0.999992 & 0.999988 & 0.999982 \\  
     20 & 0.999993 & 0.999986 & 0.999980 & 0.999969 \\  
     25 & 0.999989 & 0.999978 & 0.999968 & 0.999952 \\  
     30 & 0.999984 & 0.999968 & 0.999955 & 0.999931 \\  
     50 & 0.999958 & 0.999913 & 0.999879 & 0.999812 \\  
     70 & 0.999919 & 0.999833 & 0.999770 & 0.999638 \\  
    100 & 0.999840 & 0.999666 & 0.999548 & 0.999281 \\  
    150 & 0.999655 & 0.999271 & 0.999034 & 0.998446 \\  
    200 & 0.999408 & 0.998742 & 0.998358 & 0.997337 \\  
    250 & 0.999106 & 0.998088 & 0.997533 & 0.995980 \\  
    300 & 0.998750 & 0.997318 & 0.996573 & 0.994397 \\  
    500 & 0.996857 & 0.993223 & 0.991568 & 0.986162 \\  
    700 & 0.994325 & 0.987790 & 0.985052 & 0.975560 \\  
   1000 & 0.989563 & 0.977772 & 0.973189 & 0.956657 \\  
   1500 & 0.979675 & 0.957749 & 0.949690 & 0.920633 \\  
   2000 & 0.968032 & 0.935332 & 0.923467 & 0.882350 \\  
   2500 & 0.955170 & 0.911784 & 0.895919 & 0.843974 \\  
   3000 & 0.941459 & 0.887865 & 0.867915 & 0.806627 \\  
   5000 & 0.882570 & 0.795572 & 0.759993 & 0.675026 \\  
   7000 & 0.822962 & 0.714364 & 0.666219 & 0.572701 \\  
  10000 & 0.739435 & 0.614288 & 0.554000 & 0.460880 \\  
  15000 & 0.622323 & 0.492124 & 0.424424 & 0.342480 \\  
  20000 & 0.530963 & 0.407029 & 0.340027 & 0.270048 \\  
  25000 & 0.459732 & 0.345261 & 0.282000 & 0.221929 \\  
  30000 & 0.403520 & 0.298792 & 0.240136 & 0.187921 \\  
  50000 & 0.266133 & 0.191769 & 0.149197 & 0.115667 \\  
  70000 & 0.196376 & 0.140118 & 0.107696 & 0.083251 \\  
 100000 & 0.140107 & 0.099381 & 0.075824 & 0.058519 \\ 
 \multicolumn{1}{c}{ }& \multicolumn{4}{c}{Asymptotic} \\
 100000 & 0.142912& 0.100738 & 0.076257 & 0.058760 \\
\end{tabular}	       				   
\end{center}
\end{table}

\begin{table}
\begin{center}
\caption{The coefficients $W_4$ and $W_6$ appearing in the expansion
of the atom-atom interaction potential at small distances,
see Eq.~(\protect\ref{atom-small}).}
\label{rel-table}
\begin{tabular}{l l r@{}l r@{}l}
 &\multicolumn{1}{c}{Ref.} & \multicolumn{2}{c}{$W_4$} 
           &\multicolumn{2}{c}{$W_6$} \\
\hline
He($2\,^1\!S$)-He($2\,^1\!S$) & This work &
        3&.912\,7(5)   &  555&.86(5) \\
He($2\,^3\!S$)-He($2\,^3\!S$)  & Chen and Chung~\protect\cite{CheChu96}& 
        3&.3006   & 314&.18\protect\tablenote{Multiplied by the factor 
                   $\case{3}{2}$ to correspond to the theory of 
                   Ref.~\protect\cite{PowThi96}.} \\
                              & This work &
        3&.305\,2(5)   &  314&.44(5) \\
\end{tabular}	       				   
\end{center}
\end{table}
\begin{table}
\begin{center}
\caption{For He($2\,^1\!S$),
values of $-R^3V_{{\rm At}D}(R,\epsilon)$,
where $V_{{\rm At}D}(R,\epsilon)$ is
the atom-wall potential, for
values of $\epsilon$ corresponding
to several types of dielectric,
cols.~2--4,
and in col.~5 values of $-R^3V_{{\rm At}M}(R)$
for a perfectly conducting wall. The coefficient
$C_3(2\,^1\!S)$ is 2.67121.}
\label{glass-table-singlet}
\begin{tabular}{lcccc}
    $R$ &  $\epsilon=2.123$ & $\epsilon=2.295$ &
           $\epsilon=3.493$ & $\epsilon=\infty$\\
\hline
     10 & 0.95339 & 1.04221 & 1.47123& 2.65990 \\ 
     15 & 0.95029 & 1.03882 & 1.46644& 2.65455 \\ 
     20 & 0.94739 & 1.03564 & 1.46194& 2.64938 \\ 
     25 & 0.94463 & 1.03262 & 1.45768& 2.64439 \\ 
     30 & 0.94200 & 1.02975 & 1.45361& 2.63956 \\ 
     50 & 0.93244 & 1.01928 & 1.43883& 2.62159 \\ 
     70 & 0.92395 & 1.00999 & 1.42570& 2.60532 \\ 
    100 & 0.91253 & 0.99750 & 1.40805& 2.58320 \\ 
    150 & 0.89577 & 0.97916 & 1.38214& 2.55042 \\ 
    200 & 0.88087 & 0.96286 & 1.35913& 2.52098 \\ 
    250 & 0.86726 & 0.94797 & 1.33812& 2.49381 \\ 
    300 & 0.85463 & 0.93415 & 1.31863& 2.46833 \\ 
    500 & 0.81082 & 0.88624 & 1.25109& 2.37768 \\ 
    700 & 0.77415 & 0.84615 & 1.19462& 2.29898 \\ 
   1000 & 0.72765 & 0.79532 & 1.12305& 2.19547 \\ 
   1500 & 0.66478 & 0.72660 & 1.02637& 2.04896 \\ 
   2000 & 0.61386 & 0.67095 & 0.94810& 1.92470 \\ 
   2500 & 0.57109 & 0.62421 & 0.88236& 1.81635 \\ 
   3000 & 0.53433 & 0.58405 & 0.82587& 1.72023 \\ 
   5000 & 0.42573 & 0.46540 & 0.65887& 1.41942 \\ 
   7000 & 0.35344 & 0.38641 & 0.54755& 1.20473 \\ 
  10000 & 0.28060 & 0.30682 & 0.43519& 0.97660 \\ 
  15000 & 0.20730 & 0.22670 & 0.32187& 0.73525 \\ 
  20000 & 0.16342 & 0.17873 & 0.25390& 0.58540 \\ 
  25000 & 0.13443 & 0.14702 & 0.20893& 0.48434 \\ 
  30000 & 0.11395 & 0.12463 & 0.17715& 0.41205 \\ 
  50000 & 0.07032 & 0.07692 & 0.10938& 0.25591 \\ 
  70000 & 0.05067 & 0.05543 & 0.07883& 0.18478 \\ 
  100000 & 0.03565 & 0.03899 & 0.05545& 0.13013 \\ 
\end{tabular}	       				   
\end{center}
\end{table}
\begin{table}
\begin{center}
\caption{For He($2\,^3\!S$),
values of $-R^3V_{{\rm At}D}(R,\epsilon)$,
where $V_{{\rm At}D}(R,\epsilon)$ is
the atom-wall potential, for
values of $\epsilon$ corresponding
to several types of dielectric,
cols.~2--4,
and in col.~5 values of $-R^3V_{{\rm At}M}(R)$
for a perfectly conducting wall. The coefficient
$C_3(2\,^3\!S)$ is 1.90092.}
\label{glass-table-triplet}
\begin{tabular}{lcccc}
    $R$ &  $\epsilon=2.123$ & $\epsilon=2.295$ &
           $\epsilon=3.493$ & $\epsilon=\infty$\\
\hline
     10 & 0.67644 & 0.73946 & 1.04384& 1.88963 \\ 
     15 & 0.67336 & 0.73608 & 1.03907& 1.88428 \\ 
     20 & 0.67047 & 0.73292 & 1.03459& 1.87912 \\ 
     25 & 0.66773 & 0.72992 & 1.03036& 1.87413 \\ 
     30 & 0.66512 & 0.72707 & 1.02633& 1.86931 \\ 
     50 & 0.65566 & 0.71671 & 1.01169& 1.85142 \\ 
     70 & 0.64728 & 0.70755 & 0.99875& 1.83525 \\ 
    100 & 0.63606 & 0.69527 & 0.98140& 1.81333 \\ 
    150 & 0.61966 & 0.67733 & 0.95607& 1.78095 \\ 
    200 & 0.60516 & 0.66146 & 0.93368& 1.75197 \\ 
    250 & 0.59196 & 0.64703 & 0.91333& 1.72529 \\ 
    300 & 0.57977 & 0.63370 & 0.89454& 1.70030 \\ 
    500 & 0.53788 & 0.58789 & 0.83004& 1.61162 \\ 
    700 & 0.50336 & 0.55016 & 0.77695& 1.53484 \\ 
   1000 & 0.46046 & 0.50328 & 0.71103& 1.43436 \\ 
   1500 & 0.40433 & 0.44195 & 0.62481& 1.29413 \\ 
   2000 & 0.36074 & 0.39433 & 0.55783& 1.17825 \\ 
   2500 & 0.32560 & 0.35593 & 0.50380& 1.08034 \\ 
   3000 & 0.29654 & 0.32419 & 0.45909& 0.99640 \\ 
   5000 & 0.21739 & 0.23769 & 0.33709& 0.75412 \\ 
   7000 & 0.17046 & 0.18640 & 0.26458& 0.60138 \\ 
  10000 & 0.12784 & 0.13981 & 0.19860& 0.45715 \\ 
  15000 & 0.08946 & 0.09785 & 0.13907& 0.32318 \\ 
  20000 & 0.06848 & 0.07490 & 0.10649& 0.24852 \\ 
  25000 & 0.05536 & 0.06055 & 0.08610& 0.20136 \\ 
  30000 & 0.04641 & 0.05076 & 0.07219& 0.16904 \\ 
  50000 & 0.02810 & 0.03074 & 0.04372& 0.10257 \\ 
  70000 & 0.02012 & 0.02201 & 0.03131& 0.07350 \\ 
  100000 & 0.01411 & 0.01543 & 0.02195& 0.05154 \\ 
\end{tabular}	       				   
\end{center}
\end{table}
\clearpage
\clearpage
\clearpage
\clearpage

\end{document}